\documentclass[journal]{IEEEtran}

\usepackage{graphicx}
\usepackage{amssymb}
\usepackage{amsfonts}
\usepackage{amsmath}
\usepackage{epsfig}
\usepackage{color}
\usepackage{fancybox}
\usepackage{textcomp}
\usepackage{multirow}
\usepackage{setspa ce}
\usepackage{psfrag}
\usepackage[ruled,vlined,linesnumbered]{algorithm2e}

\usepackage{hyperref}

\usepackage{mathrsfs}

    \def\Complex{{\rm\rule[.23ex]{.03em}{1.1ex}\kern-.3em{C}}}

    \newcommand{\be}{\begin{equation}} \newcommand{\ee}{\end{equation}}
    \newcommand{\bea}{\begin{eqnarray}} \newcommand{\eea}{\end{eqnarray}}
    \newcommand{\benum}{\begin{enumerate}} \newcommand{\eenum}{\end{enumerate}}

    \newcommand{\qh}{{\bf h}}

    \newcommand{\qs}{{\bf s}}

    \newcommand{\qv}{{\bf v}}

    \newcommand{\qF}{{\bf F}}
    
    \newcommand{\qH}{{\bf H}}

    \newcommand{\tqh}{\tilde{\bf h}}

    \newcommand{\tqH}{\tilde{\bf H}}

    \newcommand{\bbC}{{\mathbb C}}

    \newcommand{\Ex}{{\sf E}}

\begin{document}

\title{Deep Learning for Massive MIMO CSI Feedback}

\author{Chao-Kai~Wen, Wan-Ting~Shih,~and~Shi~Jin
\thanks{C.-K. Wen and W.-T. Shih are with the Institute of Communications Engineering, National Sun Yat-sen University, Kaohsiung, Taiwan (e-mail: $\rm chaokai.wen@mail.nsysu.edu.tw$, $\rm sydney2317076@gmail.com$).}

\thanks{S. Jin is with the National Mobile Communications Research Laboratory, Southeast University, Nanjing 210096, P. R. China (e-mail: $\rm jinshi@seu.edu.cn$).
}

\thanks{The source code of this paper is available on
GitHub: \url{https://github.com/sydney222/Python_CsiNet}.}

%\thanks{The work of C.-K. Wen was supported by the ITRI in Hsinchu, Taiwan,
%under the project entitled ``FY104 Accelerating the development of mobile broadband services and industry-Development on emerging mobile broadband system technologies'' and the MOST of Taiwan
%under Grant MOST103-2221-E-110-029-MY3. The work of S. Jin was supported by the National Natural Science  Foundation of China under Grant 61222102 and  the International Science \& Technology Cooperation Program of China under Grant 2014DFT10300. The work of G. Wu was supported by S\&T Collaboration Project under Grant 2015DFT10170 by MOST of China.
%}
}

%\markboth{IEEE Transactions on Wireless Communications,~Vol.~XX, No.~XX, XXX~2013}%
%{Shell \MakeLowercase{\textit{et al.}}: Bare Demo of IEEEtran.cls for Journals}

\maketitle

\begin{abstract}
In frequency division duplex mode, the downlink channel state information (CSI) should be sent to the base station through feedback links so that the potential gains of a massive multiple-input multiple-output can be exhibited. However, such a transmission is hindered by excessive feedback overhead. In this letter, we use deep learning technology to
develop CsiNet, a novel CSI sensing and recovery {mechanism} that learns to effectively use channel structure from training samples.
CsiNet learns a transformation from CSI to a near-optimal number of representations (or codewords) and an inverse transformation from codewords to CSI. We perform experiments to demonstrate that CsiNet can recover CSI with significantly improved reconstruction quality compared with existing compressive sensing (CS)-based methods.
Even at excessively low compression regions where CS-based methods cannot work, CsiNet retains effective beamforming gain.
\end{abstract}


\section*{I. Introduction}

The massive multiple-input multiple-output (MIMO) system is widely regarded as a major
technology for fifth-generation wireless communication
systems. By equipping a base station (BS) with hundreds or even thousands of antennas in a centralized
\cite{Marzetta-10TW} or distributed \cite{Zhang-13JSAC} manner, such a system can substantially reduce
multiuser interference and provide a multifold increase in cell throughput.
This potential benefit is mainly obtained by exploiting channel state information (CSI) at BSs.
In current frequency division duplexity (FDD) MIMO systems (e.g., long-term evolution Release-8), the downlink CSI is acquired at the user equipment (UE) during the training period and returns to the BS through feedback links. Vector quantization or codebook-based approaches are usually adopted to reduce feedback overhead. However, the feedback quantities resulting from these approaches need to be scaled linearly with the number of transmit antennas and are prohibitive in a massive MIMO regime.

The challenge of CSI feedback in massive MIMO systems has motivated numerous studies \cite{Kuo-12WCNC,Rao-2014TSP}. These works have mainly focused on reducing feedback overhead by using the spatial and temporal correlation of CSI. In particular, correlated CSI can be transformed into an uncorrelated sparse vector in some bases; thus, one can use compressive sensing (CS) to obtain a sufficiently accurate estimate of a sparse vector from an underdetermined linear system. This concept has inspired the establishment of CSI feedback protocols based on CS \cite{Kuo-12WCNC} and distributed compressive channel estimation \cite{Rao-2014TSP}.
{The use of several algorithms, including LASSO $\ell_1$-solver \cite{Daubechies-04PAMath} and AMP \cite{Donoho-09NAS}, has also been proposed in CS.
However, these algorithms \cite{Daubechies-04PAMath,Donoho-09NAS} struggle to recover compressive CSI because they use a simple sparsity prior while their channel matrix is not perfectly but is \emph{approximately} sparse. Moreover, the changes among most adjacent elements in the channel matrix are subtle.
These properties complicate modeling their priors.
Although researchers have designed advanced algorithms (e.g., TVAL3 \cite{Li-09TVAL3} and BM3D-AMP \cite{Metzler-16TIT}) that can impose elaborate priors on reconstruction,
these algorithms do not significantly boost CSI recovery quality because hand-crafted priors remain far from practice.}

Summarily, three central problems are inherent in CS-based methods. First, they rely heavily on the assumption that channels are sparse in some bases. However, channels are not exactly sparse in any basis and may \emph{not} even have an interpretable structure. Second, CS uses random projection and does not fully exploit channel structures. Third, existing CS algorithms for signal reconstruction are often iterative approaches, {which have slow reconstruction.}
In the present study, we address the above problems using deep learning (DL).
DL attempts to mimic the human brain to accomplish a specific task by training large multilayered neural networks with vast numbers of training samples.
Our developed {\bf CSI} sensing (or encoder) and recovery (or decoder) {\bf net}work is hereafter called CsiNet.
CsiNet has the following features.
\begin{itemize}
\item \emph{Encoder}. Rather than using random projection, CsiNet learns a transformation from original channel matrices to compress representations (codewords) through training data. The algorithm is agnostic to human knowledge on channel distribution and instead directly learns to effectively use the channel structure from training data.

\item \emph{Decoder}. CsiNet learns inverse transformation from codewords to original channels. Inverse transformation is non-iterative and multiple orders of magnitude faster than iterative algorithms.
\end{itemize}

A UE uses the encoder to transform channel matrices into codewords. Once the codewords are returned to the BS, it recovers the original channel matrices by using the decoder. The methodology can be used in FDD MIMO systems as a feedback protocol. In fact, CsiNet is closely related to the autoencoder \cite[Ch.\,14]{Goodfellow-et-al-2016} in DL, which is used to learn a representation (encoding) for a set of data typically for dimensionality reduction.
Recently, several DL architectures have been proposed to reconstruct natural images from CS measurements \cite{Lohit-17ArXiv,Mousavi-17ArXiv,Yao-17ArXiv}.
Although DL exhibits state-of-the-art performance in natural-image reconstruction, whether DL
can also show its ability in wireless channel reconstruction is unclear because
this reconstruction is more sophisticated than image reconstruction.
The present work is the first to suggest a DL-based CSI reduction and recovery approach.\footnote{For an overview of applying DL to the wireless physical layer, we refer the interested readers to \cite{Wang-17ArXiv}.}
The most relevant work appears to be \cite{Shea-17ArXiv}, in which DL-based CSI encoding has been used in a closed-loop MIMO system.
Different from \cite{Shea-17ArXiv}, which has not considered CSI recovery, we show that CSI can be recovered with significantly improved reconstruction quality through DL compared with existing CS-based approaches. Even reconstructions at an excessively low compression rate retain sufficient content that allows effective beamforming gain.

\section*{II. System Model and CSI Feedback}

We consider a {simple single-cell} downlink massive MIMO system with $N_{\rm t} \gg 1$ transmit antennas at a BS and a single receiver antenna at a UE. The system is operated in OFDM over $\tilde{N}_{\rm c}$ subcarriers.
The received signal at the $n$th subcarrier is provided as follows:
\begin{equation}
    y_n = \tqh_{n}^{H} \qv_n x_n + z_n,
\end{equation}
where $\tqh_n \in \bbC^{N_{\rm t} \times 1}$, $\qv_n \in \bbC^{N_{\rm t} \times 1}$, ${x_n \in \bbC}$, and ${z_n \in \bbC}$ denote
the channel vector, precoding vector, data-bearing symbol, and additive noise of the $n$th subcarrier, respectively.
Let $\tqH = [\tqh_1 \ldots \tqh_{\tilde{N}_{\rm c}}]^H \in \bbC^{ \tilde{N}_{\rm c} \times N_{\rm t}}$
be the CSI stacked in the spatial frequency domain.
The BS can design the precoding vectors $\{\qv_n, \, n=1,\ldots, \tilde{N}_{\rm c}\}$ once it receives $\tqH$ feedback.
In the FDD system, the UE should return $\tqH$ to the BS through feedback links.
The total number of feedback parameters is $N_{\rm t} \tilde{N}_{\rm c}$, which is not allowed for limited feedback links.
Although downlink channel estimation is challenging, this topic is beyond the scope of this paper. {We
assume that perfect CSI has been acquired through pilot-based training \cite{Choi-14JSTSP} and focus on the feedback scheme.}

\begin{figure*}
    \centering
    \resizebox{6.5in}{!}{%
    \includegraphics*{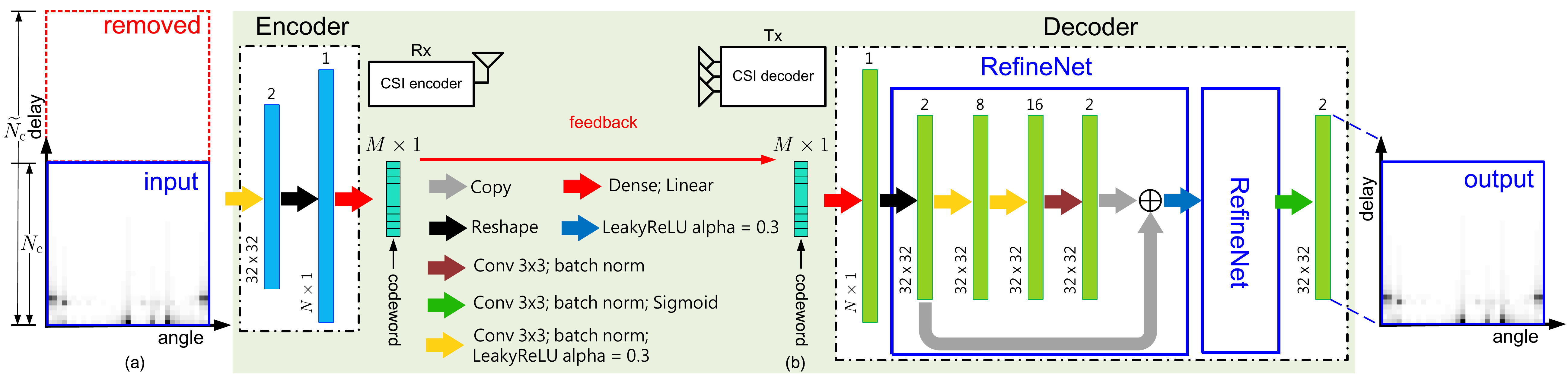} }%
    \caption{(a) Pseudo-color plot of the strength of $\qH \in \bbC^{32 \times 32}$.
    (b) Architecture of CsiNet, which includes the encoder and decoder. \label{fig:CS-CsiNet}}
\end{figure*}

To reduce feedback overhead, we propose that $\tqH$ can be sparsitied in the angular-delay domain using a 2D discrete Fourier transform (DFT) as follows:
\begin{equation} \label{eq:HtoDFTs}
    \qH =\qF_{\sf d} \tqH \qF_{\sf a}^{H},
\end{equation}
where $\qF_{\sf d}$ and $\qF_{\sf a}$ are ${\tilde{N}_{\rm c} \times \tilde{N}_{\rm c}}$ and ${N_{\rm t} \times N_{\rm t}}$ DFT matrices, respectively.
To clarify this concept, a realization of the absolute values of $\qH$ with the COST 2100 channel model \cite{Liu-12WirelessCom} is depicted in Fig.~\ref{fig:CS-CsiNet}(a). Parameterization is performed using a uniform linear array (ULA) with half-wavelength spacing in an indoor environment. The elements of $\qH$ contain only a small fraction of large components, and the other components are close to zero.
In the delay domain, only the first $N_{\rm c}$ rows of $\qH$ contain values
because the time delay between multipath arrivals lies within a limited period. Therefore, we can retain the first $N_{\rm c}$ rows of $\qH$ and remove remaining rows. By an abuse of notation, we continuously use $\qH$ to denote the ${N_{\rm c} \times N_{\rm t}}$ truncated matrix. The total number of feedback parameters can be reduced to $N = N_{\rm c} N_{\rm t}$, which remains a large number in the massive MIMO regime.

In this study, we are interested in designing the encoder
\begin{equation} \label{eq:encoder}
    \qs = f_{\sf en}(\qH),
\end{equation}
which can transform the channel matrix into an $M$-dimensional vector (codeword), where $M < N$. The data compression ratio is $\gamma = M/N$. In addition, we have to design the inverse transformation (decoder) from the codeword
to the original channel, that is,
\begin{equation} \label{eq:decoder}
    \qH = f_{\sf de}(\qs).
\end{equation}
The CSI feedback approach is as follows.
Once the channel matrix $\tqH$ is acquired at the UE side, we perform 2D DFT in \eqref{eq:HtoDFTs} to obtain the truncated matrix $\qH$ and then use the encoder \eqref{eq:encoder} to generate a codeword $\qs$. Next, $\qs$ is returned to the BS, and the BS uses the decoder \eqref{eq:decoder} to obtain $\qH$. The final channel matrix in the spatial-frequency domain can be obtained by performing inverse DFT.

\section*{III. CsiNet}

% \subsection*{A. Architecture}

We exploit the recent and popular conventional neural networks (CNNs) for the encoder and decoder they can exploit spatial local correlation by enforcing a local connectivity pattern among the neurons of adjacent layers.
The overview of the proposed DL architecture, named CsiNet, is shown in Fig. \ref{fig:CS-CsiNet}(b), in which
the values $S_1 \times S_2 \times S_3 $ denote the length, width, and number of feature maps, respectively.
The first layer of the encoder is a convolutional layer with the real and imaginary parts of $\qH$ being its input. This layer uses kernels with dimensions of ${3 \times 3}$ to generate two feature maps.
Following the convolutional layer, we reshape the feature maps into a vector and use a fully connected layer to generate the codeword $\qs$, which is a real-valued vector of size $M$. The first two layers mimic the projection of CS and serve as encoders. However, in contrast to random projections in CS, CsiNet attempts to translate the extracted feature maps into a codeword.

Once we obtain the codeword $\qs$, we use several layers (as a decoder) to map it back into the channel matrix $\qH$.
The first layer of the decoder is a fully connected layer that considers $\qs$ as input and outputs two matrices of size ${N_{\rm c} \times N_{\rm t}}$, which serve as an initial estimate of the real and imaginary parts of $\qH$.
The initial estimate is then fed into several ``RefineNet units'' that continuously refine the reconstruction.
Each RefineNet unit consists of four layers, as shown in Fig. \ref{fig:CS-CsiNet}(b).
In RefineNet unit, the first layer is the input layer.
All the remaining 3 layers use $3 \times 3$ kernels.
The second and third layers generate $8$ and $16$ feature maps, respectively, and
the final layer generates the final reconstruction of $\qH$.
Using appropriate zero padding, the feature maps produced by the three convolutional layers are set to the same size as the input channel matrix size ${N_{\rm c} \times N_{\rm t}}$.
The rectified linear unit (ReLU), ${\sf ReLU}(x) = \max(x,0)$, is used as the activation function, and we introduce batch normalization to each layer.

Two features of a RefineNet unit are as follows.
First, the output size of the RefineNet unit is equal to the channel matrix size. This concept is inspired by \cite{Lohit-17ArXiv,Mousavi-17ArXiv}.
To reduce dimensionality, nearly all conventional implementations of CNNs involve pooling layers, which is a form of down-sampling.
In contrast to conventional implementations, our target is refinement rather than dimensionality reduction.
Second, in the RefineNet unit, we introduce identity shortcut
connections that directly pass data flow to later layers.
This approach is inspired by the deep Residual Network \cite{He-16CVPR,Yao-17ArXiv}, which avoids the vanishing gradient problem caused by multiple stacked non-linear transformations.

% With the identity shortcut connections, RefineNet performs better performance than other DL architectures.

Experiments reveal that two RefineNet units produce good performance. Adding further RefineNet units does not significantly boost reconstruction quality but adds to computational complexity. Once the channel matrix has been refined by a series of RefineNet units, the channel matrix is input into the final convolutional layer, and the sigmoid function is used to scale values to the $[0, 1]$ range. {CsiNet can be extended to deal with cases involving multiple antennas at the UE by increasing the numbers of feature maps, i.e., $S_3$. We leave the exploitation of the spatial correlation across UE antennas as a topic for future studies.}

% \subsection*{B. Training}

To train CsiNet, we use end-to-end learning for all the kernel and bias values of the encoder and decoder. This training procedure differs from the two-step approach used in \cite{Yao-17ArXiv}.
The set of parameters is denoted as $\Theta = \{\Theta_{\rm en}, \Theta_{\rm de} \}$.
The input to CsiNet is $\qH_i$, and the reconstructed channel matrix is denoted by $\widehat{\qH}_i = f(\qH_i; \Theta) \triangleq f_{\rm de}( f_{\rm en}(\qH_i; \Theta_{\rm en}); \Theta_{\rm de})$ for the $i$th patch. Notably, the input and output of CsiNet are \emph{normalized} channel matrices, whose elements are scaled in the $[0, 1]$ range.
Similar to the autoencoder, CsiNet is an unsupervised learning algorithm.
The set of parameters is updated by the ADAM algorithm.
The loss function is the mean squared error (MSE), which is calculated as follows:
\begin{equation}
    L(\Theta) = \frac{1}{T} \sum_{i=1}^{T} \| f(\qs_i; \Theta) - \qH_i \|_2^{2},
\end{equation}
where the norm $\| \cdot \|_2$ is the Euclidean norm, and $T$ is the total number of samples in the training set.

\vspace{-0.25cm}
\section*{IV. Experiments}

%\subsection*{A. Implementation Details}

To generate the training and testing samples, we create two types of channel matrices through the COST 2100 channel model \cite{Liu-12WirelessCom}: 1) the indoor picocellular scenario at the $5.3$\,GHz band, and 2) the outdoor rural scenario at the $300$\,MHz band. All parameters follow their default setting in \cite{Liu-12WirelessCom}. The BS is positioned at the center of a square area with lengths of $20$ and $400$m for indoor and outdoor scenarios, respectively, whereas the UEs are randomly positioned in the square area per sample.
We use the ULA with $N_{\rm t} = 32$ antennas at the BS and ${\tilde{N}_{\rm c} = 1024}$ subcarriers.
When transforming the channel matrix into the angular-delay domain, we retain the first $32$ rows of the channel matrix. That is, $\qH$ is ${32 \times 32}$ in size.
The training, validation, and testing sets contain 100,000, 30,000, and 20,000 samples, respectively.
All testing samples are excluded from the training and validation samples.
We train several parameter sets with Glorot uniform initialization and then select the parameter set that provides minimal loss in the validation test. The epochs, learning rate, and batch size are set as $1000$, $0.001$, and $200$, respectively.

\begin{table}
\centering
\caption{NMSE in {\rm dB} and cosine similarity $\rho$.}
\label{table:result}
\begin{footnotesize}
\begin{tabular}{cl|cccc}
\hline
\multirow{2}{*}{$\gamma$} & \multicolumn{1}{c|}{\multirow{2}{*}{Methods}} & \multicolumn{2}{c}{Indoor}      & \multicolumn{2}{c}{Outdoor}    \\ \cline{3-6}
                          & \multicolumn{1}{c|}{}                         & NMSE            & $\rho$         & NMSE           & $\rho$         \\ \hline\hline
                          & LASSO                                         & -7.59           & 0.91          & -5.08          & 0.82          \\
                          & BM3D-AMP                                      & -4.33           & 0.80          & -1.33          & 0.52          \\
1/4                       & TVAL3                                         & -14.87          & 0.97          & -6.90          & 0.88          \\
                          & CS-CsiNet                                     & -11.82          & 0.96          & -6.69          & 0.87          \\
                          & CsiNet                                        & \textbf{-17.36} & \textbf{0.99} & \textbf{-8.75} & \textbf{0.91} \\ \hline
                          & LASSO                                         & -2.72           & 0.70          & -1.01          & 0.46          \\
                          & BM3D-AMP                                      & 0.26            & 0.16          & 0.55           & 0.11          \\
1/16                      & TVAL3                                         & -2.61           & 0.66          & -0.43          & 0.45          \\
                          & CS-CsiNet                                     & -6.09           & 0.87          & -2.51          & 0.66          \\
                          & CsiNet                                        & \textbf{-8.65}  & \textbf{0.93} & \textbf{-4.51} & \textbf{0.79} \\ \hline
                          & LASSO                                         & -1.03           & 0.48          & -0.24          & 0.27          \\
                          & BM3D-AMP                                      & 24.72           & 0.04          & 22.66          & 0.04          \\
1/32                      & TVAL3                                         & -0.27           & 0.33          & 0.46           & 0.28          \\
                          & CS-CsiNet                                     & -4.67           & 0.83          & -0.52          & 0.37          \\
                          & CsiNet                                        & \textbf{-6.24}  & \textbf{0.89} & \textbf{-2.81} & \textbf{0.67} \\ \hline
                          & LASSO                                         & -0.14           & 0.22          & -0.06          & 0.12          \\
                          & BM3D-AMP                                      & 0.22            & 0.04          & 25.45          & 0.03          \\
1/64                      & TVAL3                                         & 0.63            & 0.11          & 0.76           & 0.19          \\
                          & CS-CsiNet                                     & -2.46           & 0.68          & -0.22          & 0.28          \\
                          & CsiNet                                        & \textbf{-5.84}  & \textbf{0.87} & \textbf{-1.93} & \textbf{0.59} \\ \hline
\end{tabular}
\end{footnotesize}
\end{table}

\begin{figure}
    \centering
    \resizebox{3.5in}{!}{%
    \includegraphics*{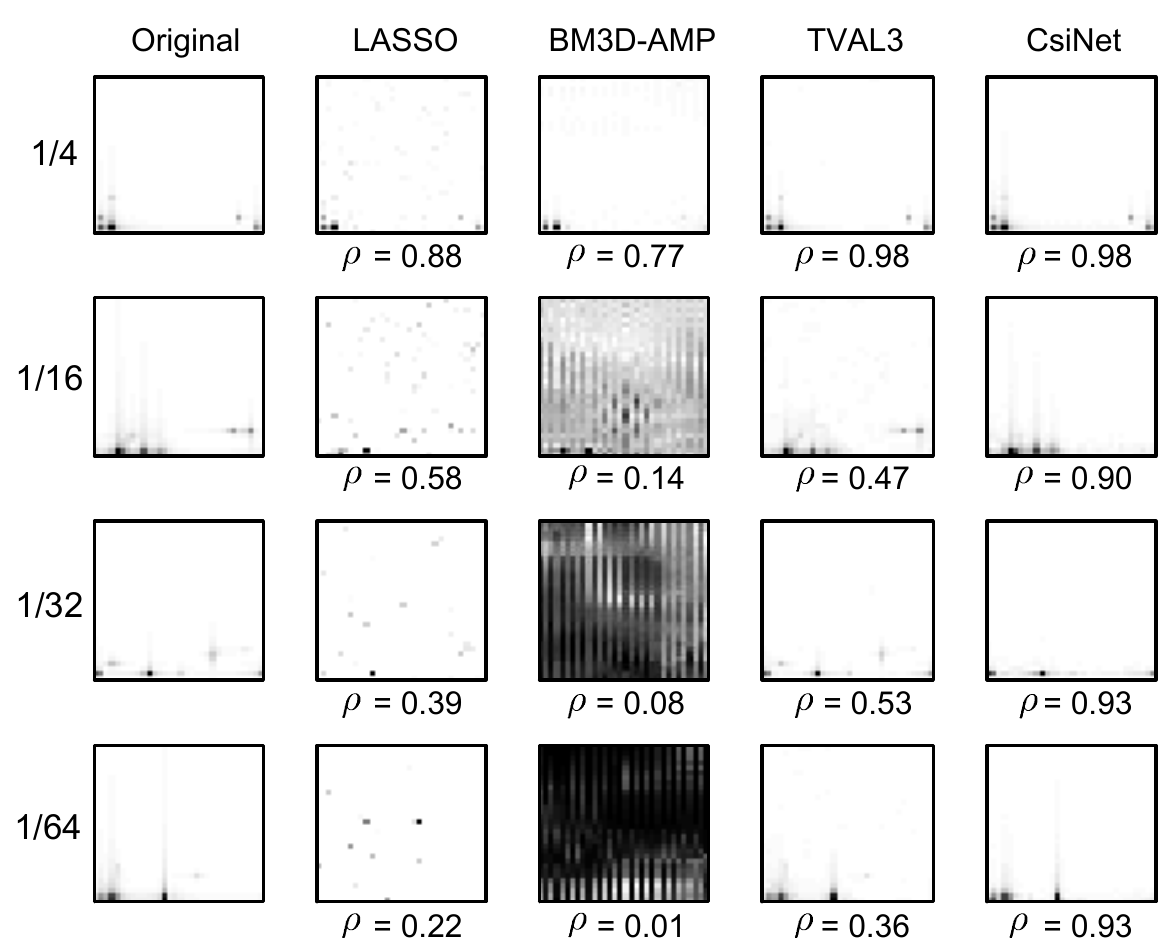} }%
    \caption{\label{fig:result}Reconstruction images for different compression ratios by different algorithms in
    indoor picocellular scenarios.}
\end{figure}
We compare CsiNet with three state-of-the-art CS-based methods, namely, LASSO $\ell_1$-solver \cite{Daubechies-04PAMath}, TVAL3 \cite{Li-09TVAL3}, and BM3D-AMP \cite{Metzler-16TIT}. In all experiments, we assume that the optimal regularization parameter of LASSO is given by an oracle.
Among these algorithms, LASSO provides the bottom-line result of the CS problem by considering only the simplest sparsity prior.
TVAL3 is a remarkably fast total variation-based recovery algorithm that considers increasingly elaborate priors. BM3D-AMP is the most accurate compressive recovery algorithm in natural-image reconstruction.
We also provide the corresponding results for CS-CsiNet, which only learns
to recover CSI from CS measurements (or random linear measurements). The architecture of CS-CsiNet is identical to that of the decoder of CsiNet.

The difference between the recovered channel $\widehat{\qH}$ and original $\qH$ is quantified by a normalized MSE, which is defined as follows:
\begin{equation}
    {\rm NMSE} = \Ex{\left\{ {\| \qH - \widehat{\qH} \|_2^2 }/{ \| \qH \|_2^2 } \right\}}.
\end{equation}
The feedback CSI serves as a beamforming vector. Let $\widehat{\tqh}_n$ be the reconstructed channel vector of the $n$th subcarrier. If $\qv_n = {\widehat{\tilde\qh}_n}/{\| \widehat{\tilde\qh}_n \|_2}$ is used as a beamforming vector, then we achieve the equivalent channel ${ \widehat{\tqh}_n^{H} \tilde\qh_n   }/{ \| \widehat{\tilde\qh}_n \|_2 }$ at the UE side. To measure the quality of the beamforming vector, we also consider the cosine similarity
\begin{equation}
    \rho = \Ex{\left\{ \frac{1}{\tilde{N}_{\rm c}}\sum_{n=1}^{\tilde{N}_{\rm c}} \frac{ | \widehat{\tqh}_n^{H} \tilde\qh_n | }{ \| \widehat{\tilde\qh}_n \|_2  \| \tilde\qh_n \|_2 } \right\}}.
\end{equation}
Notably, when evaluating NMSE and $\rho$, we recover the output of CsiNet (i.e., the normalized channel matrix) back to their original levels.

The corresponding NMSE and $\rho$ of all the concerned methods are summarized in Table \ref{table:result}, with the best results presented in bold font.
CsiNet obtains the lowest NMSE values and significantly outperforms CS-based methods at all compression ratios.
Compared with CS-CsiNet, CsiNet also provides significant gains, which are due to the sophisticated DL architecture in the encoder and decoder.
When the compression ratio is reduced to $1/16$, the CS-based methods can no longer function, whereas CsiNet and CS-CsiNet continue to perform well.
Fig. \ref{fig:result} shows some reconstruction samples at different compression ratios along with the corresponding pseudo-gray plots of the strength of $\qH$. CsiNet clearly outperforms the other algorithms.

{Furthermore, CSI recovery through CsiNet can be executed with a relatively lower overhead than that through CS-based algorithms because CsiNet requires only several layers of simple matrix-vector multiplications. Specifically, the average running times (in seconds) of LASSO, BM3D-AMP, TVAL3, and CsiNet are $0.1828$, $0.5717$, $0.3155$, and $0.0035$, respectively. CsiNet performs approximately $52$ to $163$ times faster than CS-based methods.}

{Finally, we provide some other observations. First, the DFT matrix $\qF_{\rm a}$ that is used to transform $\tqH$ from the spatial domain into the angular domain is unnecessary. Table \ref{table:space} shows
the NMSE and $\rho$ results using CsiNet with $\qF_{\rm a}$ compared to CsiNet without $\qF_{\rm a}$.
CsiNet can also exhibit good performances without employing $\qF_{\rm a}$ when retraining entire layers.
This finding demonstrates that CsiNet can learn a proper basis by itself without preprocessing the channel matrix into the angular domain and thus implies that CsiNet can be applied in other antenna configurations. Second, angular (or spatial) resolution increases with the number of antennas at the BS. The corresponding NMSE and $\rho$ of all the concerned methods when $N_{\rm t} = 16$, $32$, and $48$ are summarized in Table \ref{table:antenna}. The reconstruction performances of all the algorithms improve because $\qH$ becomes sparser.
CsiNet can be significantly improved because it is more capable of exploiting subtle changes among adjacent elements than CS-based methods.}

\begin{table}
\centering
\caption{The comparison of the  spatial domain and angular domain.}
\label{table:space}
\begin{footnotesize}
\begin{tabular}{cl|cccc}
\hline
\multirow{2}{*}{$\gamma$} & \multicolumn{1}{c|}{\multirow{2}{*}{Domain}} & \multicolumn{2}{c}{Indoor} & \multicolumn{2}{c}{Outdoor} \\ \cline{3-6}
                          & \multicolumn{1}{c|}{}                         & NMSE          & $\rho$      & NMSE          & $\rho$       \\ \hline\hline
\multirow{2}{*}{1/4}      & Spatial (without $\qF_{\rm a}$)               & -24.57      & 1.00     & -9.42       & 0.92      \\
                          & Angular (with $\qF_{\rm a}$)                  & -17.36      & 0.99     & -8.75       & 0.91      \\ \hline
\multirow{2}{*}{1/16}     & Spatial (without $\qF_{\rm a}$)               & -9.20       & 0.94     & -4.14       & 0.77      \\
                          & Angular (with $\qF_{\rm a}$)                  & -8.65       & 0.93     & -4.51       & 0.79      \\ \hline
\multirow{2}{*}{1/32}     & Spatial (without $\qF_{\rm a}$)               & -8.77       & 0.93     & -2.96       & 0.69      \\
                          & Angular (with $\qF_{\rm a}$)                  & -6.24       & 0.89     & -2.81       & 0.67      \\ \hline
\multirow{2}{*}{1/64}     & Spatial (without $\qF_{\rm a}$)               & -5.83       & 0.86     & -1.78       & 0.56      \\
                          & Angular (with $\qF_{\rm a}$)                  & -5.84       & 0.87     & -1.93       & 0.59      \\ \hline
\end{tabular}
\end{footnotesize}
\end{table}

\begin{table}[]
\centering
\caption{NMSE (dB) and $\rho$ for different $N_{\rm t}$ in outdoor rural scenarios.}
\label{table:antenna}
\begin{footnotesize}
\begin{tabular}{cl|cccccc}
\hline
\multirow{2}{*}{$\gamma$} & \multicolumn{1}{c|}{\multirow{2}{*}{Methods}} & \multicolumn{2}{c}{$N_{\rm t} = 16$} & \multicolumn{2}{c}{$N_{\rm t} = 32$} & \multicolumn{2}{c}{$N_{\rm t} = 48$} \\ \cline{3-8}
                          & \multicolumn{1}{c|}{}                         & NMSE              & $\rho$            & NMSE              & $\rho$            & NMSE               & $\rho$           \\ \hline\hline
\multirow{4}{*}{1/4}      & LASSO                                         & -4.55           & 0.80           & -5.08           & 0.82           & -5.28            & 0.83          \\
                          & BM3D-AMP                                      & -1.06           & 0.47           & -1.33           & 0.52           & -1.61            & 0.62          \\
                          & TVAL3                                         & -3.87           & 0.77           & -6.90           & 0.88           & -6.09            & 0.85          \\
                          & CsiNet                                        & \textbf{-6.13}  & \textbf{0.85}  & \textbf{-8.75}  & \textbf{0.91}  & \textbf{-12.38}  & \textbf{0.94} \\ \hline
\multirow{4}{*}{1/16}     & LASSO                                         & -0.65           & 0.44           & -1.01           & 0.46           & -1.23            & 0.51          \\
                          & BM3D-AMP                                      & 1.92            & 0.27           & 0.55            & 0.11           & 0.35             & 0.23          \\
                          & TVAL3                                         & 0.03            & 0.40           & -0.43           & 0.45           & -0.79            & 0.50          \\
                          & CsiNet                                        & \textbf{-3.44}  & \textbf{0.74}  & \textbf{-3.34}  & \textbf{0.72}  & \textbf{-5.54}   & \textbf{0.83} \\ \hline
\multirow{4}{*}{1/32}     & LASSO                                         & -0.13           & 0.27           & -0.24           & 0.27           & -0.38            & 0.34          \\
                          & BM3D-AMP                                      & 21.53           & 0.23           & 22.66           & 0.04           & 23.64            & 0.13          \\
                          & TVAL3                                         & 0.65            & 0.28           & 0.46            & 0.28           & 0.28             & 0.31          \\
                          & CsiNet                                        & \textbf{-2.30}  & \textbf{0.65}  & \textbf{-2.81}  & \textbf{0.67}  & \textbf{-3.76}   & \textbf{0.74} \\ \hline
\multirow{4}{*}{1/64}     & LASSO                                         & -0.06           & 0.12           & -0.06           & 0.12           & -0.057           & 0.16          \\
                          & BM3D-AMP                                      & 23.26           & 0.04           & 25.45           & 0.03           & 26.78            & 0.13          \\
                          & TVAL3                                         & 1.02            & 0.23           & 0.76            & 0.19           & 0.72             & 0.18          \\
                          & CsiNet                                        & \textbf{-1.24}  & \textbf{0.48}  & \textbf{-1.93}  & \textbf{0.58}  & \textbf{-2.74}   & \textbf{0.67} \\ \hline
\end{tabular}
\end{footnotesize}
\end{table}

\section*{V. Conclusion}
We used DL in CsiNet, a novel CSI sensing and recovery {mechanism}.
CsiNet performed well at low compression ratios and reduced time complexity.
We believe that its reconstruction quality can be further improved by applying advance DL technology, and we hope this study encourages future research in this direction.

\vspace{-0.38cm}
{\renewcommand{\baselinestretch}{1.1}
\begin{footnotesize}
\bibliographystyle{IEEEtran}


\end{footnotesize}}

\end{document}